# Synthesis of Large-Area WS$_2$ monolayers with Exceptional Photoluminescence

*Kathleen M. McCreary\*, Aubrey T. Hanbicki, Glenn G. Jernigan, James C. Culbertson, Berend T. Jonker*

*Naval Research Laboratory,* Washington DC 20375, USA

Monolayer WS$_2$ offers great promise for use in optical devices due to its direct bandgap and high photoluminescence intensity. While fundamental investigations can be performed on exfoliated material, large-area and high quality materials are essential for implementation of technological applications. In this work, we synthesize monolayer WS$_2$ under various controlled conditions and characterize the films using photoluminescence, Raman and x-ray photoelectron spectroscopies. We demonstrate that the introduction of hydrogen to the argon carrier gas dramatically improves the optical quality and increases the growth area of WS$_2$, resulting in films exhibiting mm$^2$ coverage. The addition of hydrogen more effectively reduces the WO$_3$ precursor and protects against oxidative etching of the synthesized monolayers. The stoichiometric WS$_2$ monolayers synthesized using Ar+H$_2$ carrier gas exhibit superior optical characteristics, with photoluminescence emission full width half maximum values below 40 meV and emission intensities nearly an order of magnitude higher than films synthesized in a pure Ar environment.

1. Introduction

Single- to few-monolayer transition metal dichalcogenides (TMDs) hold promise for technological applications in a variety of areas including photodetection,[1] flexible electronics,[2,3] and chemical sensing.[4] While $MoS_2$ has received the most attention due to its relative ease of mechanical exfoliation,[5–11] the closely related $WS_2$,[12–14] $MoSe_2$,[15,16] and $WSe_2$[17,18] are quickly gaining notice. Recently, $WS_2$ demonstrated superior optical properties compared to $MoS_2$ as measured by luminescent quantum efficiency and linewidths.[12,19,20] In addition, $WS_2$ has larger spin orbit coupling,[21] suggesting $WS_2$ may exhibit larger band edge spin splittings and stronger magnetic field effects for optoelectronic and spintronic functionalities.

Weak inter-layer van der Waals bonding enables single layers of TMD materials to be easily isolated using mechanical exfoliation in order to probe fundamental optical, electronic, and spintronic properties.[22] Unfortunately, deposited flakes tend to be randomly positioned and irregularly shaped, posing significant challenges for integration into technological applications. A cost effective and reliable means to achieve large-area, high quality materials is a foundational step for the incorporation of monolayer $WS_2$ into future technologies.

Chemical vapor deposition (CVD) has proven successful in the synthesis of wafer scale monolayer graphene material[23,24] and is showing promise in $WS_2$ synthesis.[19,20,25–33] CVD synthesis of monolayer TMD materials can be performed in a quartz tube furnace under the flow of an inert gas. In the case of $WS_2$, $WO_3$ and S precursors are used. The sulfur vapor partially reduces the $WO_3$ at elevated temperatures to form a volatile $WO_{3-x}$ species, which absorbs onto the growth substrate and subsequently reacts with sulfur to produce $WS_2$. While several independent research groups have reported successful $WS_2$ synthesis using $WO_3$ and S precursors, specific details of the procedure (precursor amount, growth substrate, growth pressure, temperature, gases, flow rates, etc.) can vary widely from lab to lab. Subsequently, properties such as growth morphology, luminescence yield, and Raman spectra exhibit considerable variations.

To better understand and improve the quality of CVD synthesized monolayer $WS_2$, it is important to investigate materials grown under various conditions in a single, well-controlled CVD system. In this work, we demonstrate that small modifications to temperature and carrier gas have substantial effects on the resulting coverage, continuity, and quality of CVD synthesized $WS_2$ monolayers. By optimizing the growth conditions, we achieve large-area (~mm$^2$) monolayer $WS_2$ with superior optical characteristics. Additionally, by introducing hydrogen gas during synthesis, we are able to prevent the oxidation of $WS_2$, which may also be relevant to synthesis of other TMD monolayers and van der Waals heterostructures.[34]

## 2. Results and Discussion

Synthesis of monolayer $WS_2$ is performed in a quartz tube furnace, as shown in Figure 1a. At the center of the furnace is positioned a quartz boat containing ~1g of $WO_3$ powder. Two $Si/SiO_2$ (275 nm) wafers are positioned face-down, directly above the precursor. The upstream wafer contains perylene-3,4,9,10-tetracarboxylic acid tetrapotassium salt (PTAS) seeding molecules, while the downstream substrate is untreated. The hexagonal PTAS molecules are carried downstream to the untreated substrate and promote lateral growth of the TMD materials.[35,36] While $WS_2$ growth occurs on both substrates, the downstream (untreated) substrate is the focus of this manuscript. Additional details regarding PTAS preparation and use can be found in the supplementary information. Sulfur is placed upstream, outside the furnace heating zone. It should be noted that great care is taken to position the $WO_3$ precursor, substrates, and the sulfur source at identical positions for each growth, as the positioning may affect the growth dynamics and resulting film quality.[27] Prior to synthesis, a pump-flush procedure is performed in the quartz tube to promote a uniform initial environment that aids in the reproducibility. The quartz tube is evacuated to 100 mTorr then refilled with Ar gas until atmospheric pressure is reached. This cycle is repeated twice before a continuous flow of Ar is utilized for growth. All recipes are performed at atmospheric pressure.

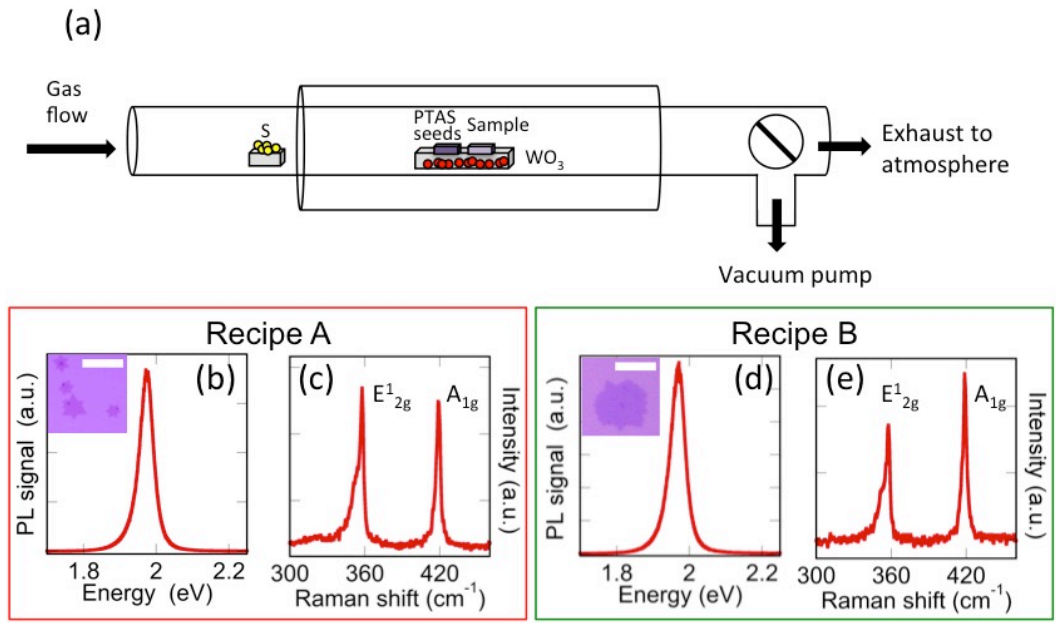

**Figure 1**: (a) Schematic of the quartz tube furnace used for $WS_2$ monolayer synthesis. (b) Photoluminescence and (c) Raman spectra confirm the identity and monolayer nature of $WS_2$ synthesized under recipe A. An optical microscope image (inset of b) displays isolated, monolayer growth with star-like configuration. (d) Photoluminescence and (e) Raman spectra of monolayer $WS_2$ synthesized with recipe B. The inset of (d) displays an optical image of the irregular, isolated $WS_2$ monolayer growth. Scale bars are 10 μm.

The initial growth (recipe A) is performed under continuous 100 sccm argon flow. The furnace temperature is quickly ramped to 625°C at a rate of 20°C/min followed by a 10°C/min ramp to 825°C. At this point, the temperature is held constant for 10 minutes and then allowed to cool. This procedure produces isolated islands exhibiting lateral dimensions of several μm, often with star-like orientations (inset of Figure 1b). Small areas of thicker growth are observed on some $WS_2$ islands. Similar features have been previously observed in synthesized TMD materials.[11,37] This characteristic growth of isolated WS$_2$ islands occurs across several mm of the growth substrate and has been reproduced in multiple growth runs.

Photoluminescence (PL) and Raman spectroscopy confirm the identity and monolayer nature of the synthesized crystals. Raman spectroscopy and PL measurements are performed at room temperature in air using 488 nm laser excitation with intensity below 200 µW for a <1µm diameter spot size, to prevent sample damage. Photoluminescence measurements (Figure 1b) reveal a sharp emission peak with maximum intensity achieved at 1.97 eV in the investigated range from 1.55 eV to 2.50 eV. The peak energy and observation of a single, intense emission peak indicates monolayer WS$_2$ material. Additional layers (not shown) dramatically decrease the emission intensity and are accompanied by a low energy peak, caused by the transition to indirect bandgap semiconductor. The Raman spectroscopy measurements (Figure 1c) detect the in-plane and out-of plane phonon modes, $E^1_{2g}$ and $A_{1g}$, at 357.5 cm$^{-1}$ and 419 cm$^{-1}$, respectively. Additionally, the longitudinal acoustic mode at the M point, LA(M), is evident at ~350cm$^{-1}$, forming a low wavenumber shoulder on the $E^1_{2g}$ peak. The peak separation ($\Delta k$) between $E^1_{2g}$ and $A_{1g}$ is commonly used to identify the layer number of WS$_2$, with reported $\Delta k$ values of ~62 cm$^{-1}$ for monolayer and ~64 cm$^{-1}$ for bilayer.[37] We measure a peak separation of 61.5 cm$^{-1}$, confirming monolayer WS$_2$ synthesis.

It is instructive to investigate the effect of temperature, as a wide range of values have been reported for WS$_2$ synthesis. Upon increasing the growth temperature to 875°C (recipe B), the resulting WS$_2$ islands exhibit a moderate increase in lateral size, and the shape becomes more irregular (inset of Figure 1d). Equilateral triangular growth is suggestive of single crystallinity, while star-like structures (as seen in inset of Figure 1b) are produced from several rotationally symmetric grains.[38] Hence, the observed change in Fig 1d may indicate additional grain boundaries or increased defect density. Regardless of the differing growth features, the PL again displays a single sharp emission peak centered at 1.97 eV (Figure 1d). Furthermore, the dominant $E^1_{2g}$ and $A_{1g}$ Raman peaks remain at 357.5 cm$^{-1}$ and 419 cm$^{-1}$ (Figure 1e), resulting in a Δk of 61.5 cm$^{-1}$ and indicative of single monolayer growth. Growth at higher temperatures (975°C in Ar flow) dramatically reduces WS$_2$ growth, resulting in very little substrate coverage. This observation suggests that conditions at elevated temperatures may be detrimental to WS$_2$, and will be discussed further in subsequent sections.

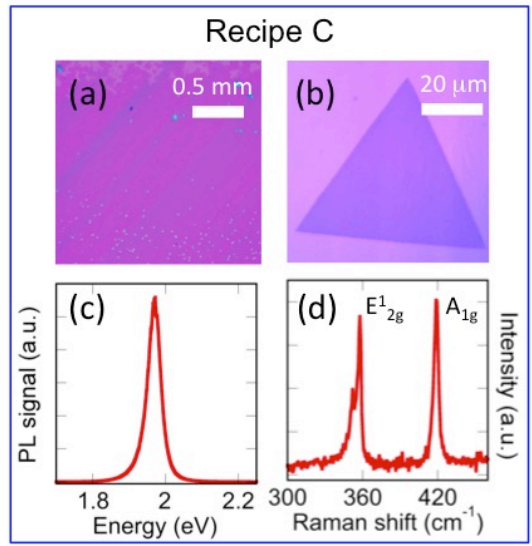

**Figure 2**: Characterization of WS$_2$ synthesized using recipe C: (a) The optical image displays large areas (~1 mm × 1 mm) of continuous monolayer WS$_2$ (purple). The edge of the film is visible at the top of the image, where areas of bare substrate are evident (pink). Very few sites of multilayer growth are present (yellow/blue). (b) An image of an isolated, equilateral WS$_2$ monolayer. (c) Photoluminescence and (d) Raman spectra, confirming monolayer WS$_2$.

The effects of carrier gas are summarized in Figure 2. Once the growth temperature of 825°C is reached, 10 sccm hydrogen is added to the 100 sccm Ar flow, and continues throughout the completion of the recipe (recipe C). The introduction of hydrogen produces substantially different growth characteristics compared to the isolated islands synthesized in recipes A and B.

WS$_2$ islands now coalesce to form nearly continuous films on the mm scale (Figure 2a). While a small fraction of multilayer growth is visually evident, most likely at grain boundaries or nucleation sites, the majority of the film is composed of monolayer material. Continuous monolayer growth on the mm scale is typical of samples synthesized using recipe C. Figure 2b shows an individual WS$_2$ crystal, at the edge of the coalesced film. Triangular growth is observed with lateral dimension larger than 50 μm, nearly an order of magnitude larger than isolated WS$_2$ islands synthesized in pure Argon at an equivalent temperature. Even though recipe C results in dramatically different growth morphology, PL and Raman spectra exhibit characteristics similar to previous recipes (Figure 1b-1d), with a single sharp PL emission at ~1.97 eV (Figure 2c) and E$^1_{2g}$ and A$_{1g}$ Raman peaks measured at 357.5 cm$^{-1}$ and 419 cm$^{-1}$ (Figure 2d).

The monolayer films synthesized using procedures A, B and C just described are further investigated by obtaining multiple Raman and PL measurements across large areas of the growth wafer. On each wafer, 5 measurement sites have been randomly selected, with each location more than 500 μm away from the closest neighbor. Additionally, only monolayer regions substantially larger than the laser spot size and free of visible multilayer growth were selected. Therefore, any measured differences are attributed to WS$_2$ quality as opposed to sample size, edge effects, or overlayer growth. While photoluminescence measurements resulted in qualitatively similar spectra, it is immediately apparent in Figure 3a, the emission intensity is extremely sensitive to the growth recipe. Samples synthesized using recipe C consistently result in the highest PL intensity, with values ranging from 94,300 to 134,300 counts/second. Recipes A and B have considerably lower intensities, with sample A ranging from 5,700 to to 28,400 counts/sec and B from 7,800 to 14,000 counts/sec (Figure 3b).

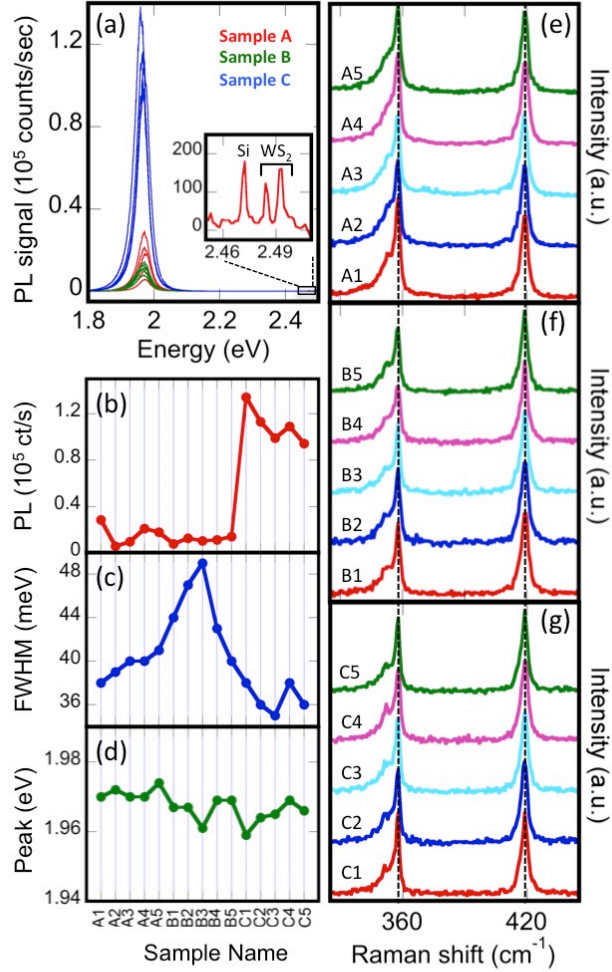

**Figure 3**: Comparisons of monolayer $WS_2$ synthesized under different recipes. (a) $WS_2$ synthesized using recipe C repeatedly results in the highest intensity PL emission. The Raman peaks from $WS_2$ and the Si substrate are also measureable (inset). The large PL/Raman ratio demonstrates the high luminescent efficiency. Values for (b) PL intensity, (c) full width at half maximum, and (d) PL peak position are shown for multiple randomly selected monolayers. (e-g) The corresponding Raman spectra are displayed. The $E^1_{2g}$ and $A_{1g}$ Raman peaks are observed at 357.5 cm$^{-1}$ and 419 cm$^{-1}$ for all $WS_2$, as indicated by dashed vertical lines.

In addition to the dominant PL emission peak, two Raman peaks ($E^1_{2g}$, $A_{1g}$) plus that of the Si substrate are evident at higher energy (inset of Figure 3a). The relation between PL and Raman intensity has been utilized as a metric to gauge optical quality of TMD materials and determine intrinsic luminescence quantum efficiency.[6] Higher PL/Raman ratios signify higher optical quality. Measurements for monolayer $MoS_2$ report a moderate PL/Raman ratio of 10 or less.[6,38–40] Recent studies of monolayer $WS_2$ report a wide range of luminescence quantum efficiencies, ranging from 1-10 to several hundred.[19,20,41] Our $WS_2$ monolayers synthesized with recipes A and B demonstrate PL/Raman ratios ranging from 20 to 120, whereas $WS_2$ synthesized under recipe C demonstrates ratios above 450 and as high as 650.

A further gauge of sample quality is the PL FWHM, with smaller FWHM indicative of higher optical quality. While all recipes exhibit FWHM below 50 meV, as previously reported for high quality $WS_2$ monolayers, it is clear that recipe C demonstrates the lowest FWHM (Figure 3c), with values as low as 35 meV corresponding to the high PL intensities measured. Conversely, samples synthesized using recipe B exhibit the highest FWHM values, consistent with the low crystalline order observed in optical images. Although substantial variations are observed in PL intensity and FWHM, the PL peak position only minor deviations from 1.97 eV, as shown in Figure 3d.

The corresponding Raman spectra measured at each site are plotted for the various samples (Figure 3e-g). Independent of recipe, the $E^1_{2g}$ and $A_{1g}$ peaks remain constant at 357.5 cm$^{-1}$ and 419 cm$^{-1}$ for all $WS_2$, as indicated by the dashed line provided as a guide to the eye. While the peak positions exhibit little variation, we do observe small modifications in the relative intensities between $E^1_{2g}$ and $A_{1g}$. Samples synthesized under recipe A tend to have a slightly higher $E^1_{2g}$ intensity, with the ratio $E^1_{2g}/A_{1g}$ ranging from 1.0 to 1.2. Conversely the $A_{1g}$ intensity is higher for recipe B, with $E^1_{2g} / A_{1g}$ from 0.7 to 0.9. The samples synthesized with recipe C exhibit nearly equal intensities with $E^1_{2g} / A_{1g}$ ~0.9 to 1.0. While moderate changes in the Raman intensity ratio are known to indicate a change in layer number for exfoliated $WS_2$,[37,42] we exclude this as the source of variation in our synthesized materials, as no increase in peak separation is observed, and independent techniques (optical microscopy, atomic force microscopy, PL) show no sizeable multilayer growth in the surveyed areas. Alternatively, minor intensity variations in $E^1_{2g}$ and $A_{1g}$ peaks of monolayer $MoS_2$ indicate differences in electronic doping levels or strain.[43] We hypothesize that the various recipes result in differences in the sulfur content of the films, which may affect both doping and local strain and contribute to the observed intensity variations, although additional investigations are necessary.

To further understand the relative differences in sample characteristics resulting from these growth processes, X-ray photoelectron spectroscopy (XPS) is used to determine the chemical composition and stoichiometry of our $WS_2$ films. Figure 4 shows the XPS spectra of the tungsten and sulfur core levels in the $WO_3$ precursor as well as $WS_2$ monolayers synthesized using recipes A, B, and C. The $WO_3$ precursor powder (Figure 4a, red line) is well fit by three peaks at binding energies 35.3 eV, 37.4 eV, and 41.3 eV corresponding to $W4f_{7/2}$, $W4f_{5/2}$, and $W5p_{3/2}$ core energy levels, respectively. The binding energies indicate the valence state of the material and are consistent with a (6+) valence state,[44,45] as expected for oxidized tungsten. A decrease in valence due to S- rather than O-bonding will be indicated by a shift in the W4f and W5p core levels to lower energies. The (6+) valence remains the dominant contribution in samples fabricated using both recipe A (Figure 4b) and recipe B (Figure 4c), indicating a large amount of $WO_3$ in these synthesized monolayers. In both samples, two additional sets of peaks are present, demonstrating the presence of partially reduced $WO_3$ (purple lines), as well as oxygen-free tungsten in the (4+) valence state (green/yellow lines), as expected for $WS_2$. The sulfur S2p peak in samples A and B (Figure 4f,g) consists of a single doublet corresponding to S-W bonding, and confirms the presence of $WS_2$. The considerably high oxygen content in these samples is surprising however, particularly in light of the photoluminescence and Raman characteristics that are comparable to high quality $WS_2$. This highlights the importance of sensitive chemical analysis techniques, such as XPS, in the rapidly progressing field of 2D materials synthesis.

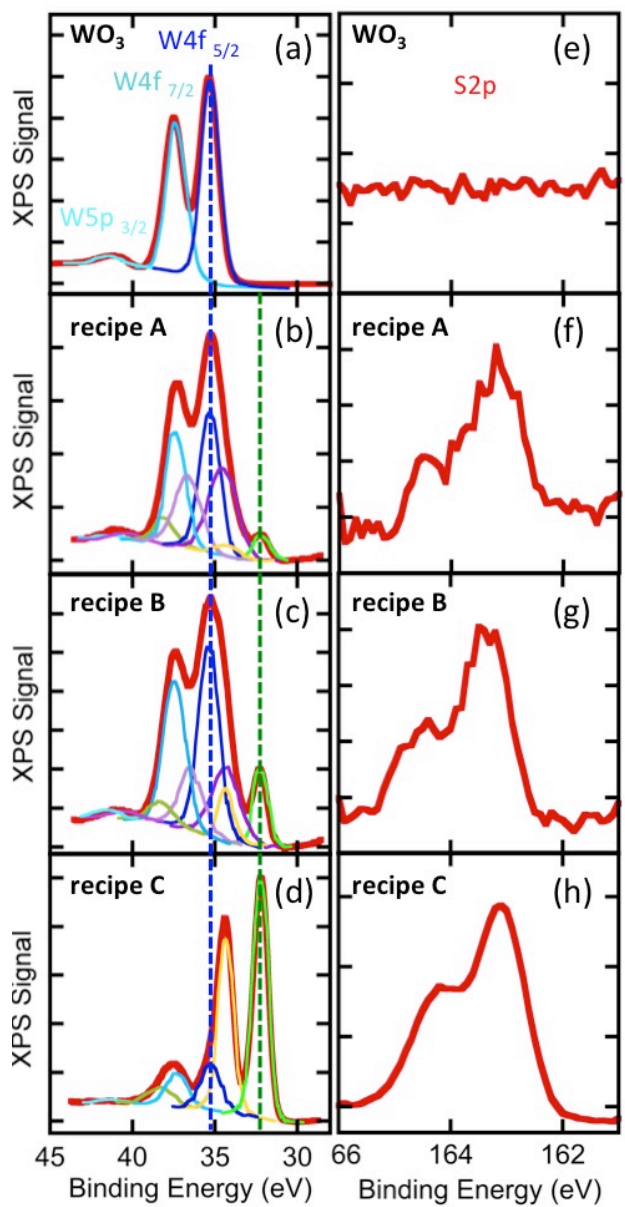

**Figure 4**: Chemical analysis using XPS. (a-d) Spectra of the tungsten core levels. (a) The $WO_3$ precursor powder (red line) is fit by three peaks (blue) corresponding to $W4f_{7/2}$, $W4f_{5/2}$, and $W5p_{3/2}$ core levels. The binding energies indicate a (6+) valence state, as expected for oxidized tungsten. (b-c) Spectra from recipes A and B exhibit $WO_3$ peaks as well as two additional sets of peaks. The additional peaks signify the presence of partially reduced $WO_3$ (purple lines), and oxygen-free tungsten in the (4+) valence state (green/yellow lines). (d) For recipe C, oxygen-free tungsten peaks dominate. Partially reduced $WO_3$ is no longer present and $WO_3$ is considerably reduced. As a guide to the eye, dashed lines indicate the $W4f_{5/2}$ binding energy in the 6+ (blue) and 4+ valence state (green). (e-h) XPS spectra of sulfur core levels. (e) No sulfur is present in the WO$_3$ precursor. (f-h) The S2p doublet confirms the presence of WS$_2$ for samples synthesized using recipes A, B, and C.

The addition of hydrogen in recipe C results in a noticeable change in the chemical composition. In the tungsten core levels, the (4+) valence state becomes the dominant contribution (Figure 4d). Peaks indicative of partially reduced WO$_3$ are no longer present, and the WO$_3$ (6+ valence) is considerably reduced. A strong S2p doublet is observed, as evident in Figure 4h. The improved sulfur signal is primarily due to the increased growth area present in these samples. Analysis of the S2p doublet and W4f peaks associated with WS$_2$ (the (4+) valence state) provide a S:W ratio of 2, indicating stoichiometric WS$_2$ synthesis for samples produced using recipe C. On the contrary, samples synthesized using recipes A and B exhibit a large sulfur deficiency, having only 45% sulfur content. It is clear that the addition of hydrogen during synthesis leads to more efficient reduction of the WO$_3$ precursor and conversion to WS$_2$. This has the multiple benefit of resulting in larger synthesis areas, improved chemical composition, and significant improvements in the optical properties.

To further investigate the effects of pure Ar and Ar/H$_2$ environments on WS$_2$, we performed a post-growth anneal procedure on WS$_2$ monolayers synthesized using procedure C. An as-grown WS$_2$ sample (Figure 5a) is returned to the furnace and exposed to the precise conditions of Recipe A (825°C, Ar=100 sccm), with the exception that no WO$_3$ precursor is present. As is evident from the optical image of the same area before- and after-anneal, the majority of the WS$_2$ on the substrate is no longer present following this procedure (Figure 5a). While the thermal stability of a WS$_2$ monolayer has yet to be determined, bulk WS$_2$ is stable up to 1250 °C, indicating that thermal decomposition is unlikely. Alternatively, ultrathin layers of TMD materials can succumb to oxidative etching at temperatures below 400 °C.[46,47] While a careful pump-purge cycle is performed prior to annealing, oxygen contamination may potentially enter during the flow of Ar and subsequent oxidative etching could cause the removal of $WS_2$. This post-growth anneal demonstrates that a pure Ar environment under atmospheric condition is not conducive to high quality $WS_2$ monolayers. While $WS_2$ is being formed through the reaction between S and $WO_3$ precursors, it is simultaneously being oxidized/ etched. As we observed for highly elevated temperatures (975 °C), the oxidative etching can dominate, resulting in no $WS_2$ by the conclusion of the synthesis process.

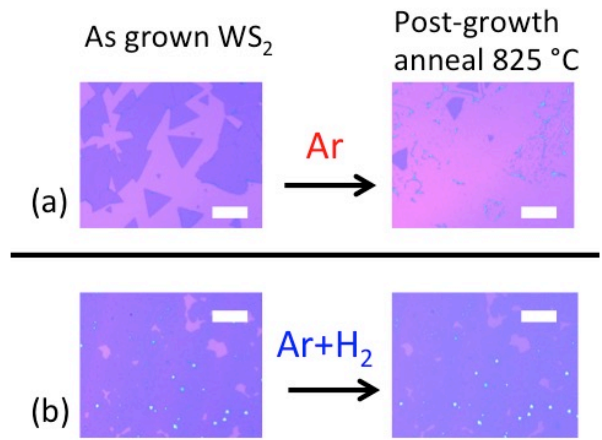

**Figure 5**: Optical images of the same sample area before-anneal and after-anneal. (a) Exposing as-grown samples to a post-growth anneal at 825°C in pure argon results in the removal of $WS_2$ monolayers. (b) The addition of hydrogen to the anneal procedure prevents damage to monolayer materials and inhibits oxidative etching of the materials. The scale bar in all images is 100 μm.

In contrast, when the post growth anneal procedure is repeated on a second as-grown sample in a flow of 10 sccm of $H_2$ and 100 sccm Ar, no oxidative etching occurs, as evident by comparing the before- and after-anneal images in Figure 5b. The $WS_2$ exhibits no change following the anneal procedure, indicating that hydrogen is inhibiting the presence of destructive oxygen. During the growth of $WS_2$, the introduction of hydrogen serves multiple purposes: it aids in the formation of $WS_2$ by reducing the precursor, leading to larger area growth of monolayer material, and it protects against oxygen damage.

## 3. Conclusion

In conclusion, we have demonstrated synthesis of monolayer $WS_2$ under various conditions. Optical characterization using photoluminescence and Raman spectroscopy of the various $WS_2$ samples show that the addition of hydrogen into the flow stream produces significantly higher PL intensity and narrower linewidths. XPS studies demonstrate that this is due to the formation of high-quality, low oxygen content monolayers achieved only when hydrogen is introduced during synthesis. The addition of hydrogen more effectively reduces the $WO_3$ precursor, resulting in monolayer films exhibiting $mm^2$ lateral dimensions, while simultaneously protecting against oxygen damage.

**Acknowledgements**
Core programs at NRL and the NRL Nanoscience Institute supported this work. This work was supported in part by the Air Force Office of Scientific Research under contract number F4GGA24233G001. This research was performed while K.M.M held a National Research Council Research Associateship Award at NRL. The authors acknowledge use of facilities in the NRL Nanoscience Institute and thank David Zapotok and Dean St. Amand for technical support.

**Author Contributions**
K.M. performed the CVD synthesis of samples. G.J. performed XPS measurements. K.M., A.H., and J.C. performed optical characterization. All authors discussed the results and contributed to the manuscript.

**Additional Information**
**Supplementary Information**: Details concerning PTAS preparation, use, and impact on $WS_2$ synthesis are available at http://www.nature.com/srep

**Competing financial interests:** The authors declare no competing financial interests.

# Supporting Information

Title: **Synthesis of Large-Area WS₂ monolayers with Exceptional Photoluminescence**

*Kathleen M. McCreary\*, Aubrey T. Hanbicki, Glenn G. Jernigan, James C. Culbertson, Berend T. Jonker*

We utilize Perylene-3,4,9,10-tetracarboxylic acid tetrapotassium salt (PTAS) to promote the growth of $WS_2$ on $Si/SiO_2$ (275 nm) substrates.[1] A ~1mM liquid solution is established by dissolving 17 mg PTAS in 30 mL water. (PTAS molecular weight = 580.7 g/mol). Approximately 30 mL of the PTAS solution is pipetted onto a clean $Si/SiO_2$ substrate, wetting the surface. The substrate is subsequently heated to 85 °C on a hotplate in ambient conditions until the water is fully evaporated. The PTAS sample is then loaded into the quartz tube face-down, directly above the $WO_3$ precursor and upstream of the bare growth substrate. During CVD synthesis, PTAS seeds are carried downstream to stimulate the growth of monolayer $WS_2$. XPS confirms the presence of a small amount of PTAS on the downstream growth substrate following all synthesis recipes (Recipe A, B, and C discussed in the main text). Small variations in concentration are measured, with samples synthesized using recipe A exhibiting the largest concentration of PTAS and recipe C the smallest.

Synthesis of $WS_2$ in the absence of PTAS seeds is also investigated. In this case, two bare $Si/SiO_2$ (275 nm) substrates are loaded face-down, directly above the $WO_3$ precursor. Synthesis conditions identical to recipes A and B result in no growth on either substrate. Synthesis conditions identical to recipe C result in $WS_2$ growth on both the upstream (Figure S1) and downstream (Figure S2) substrates. Although, in the absence of PTAS seeds, the morphology and uniformity are considerably different than when PTAS is utilized. Only isolated island growth, having lateral dimensions up to tens of mm, is observed. The synthesized islands are commonly a rounded-triangle shape, with a large amount of multilayer WS$_2$ in addition to monolayer growth. While monolayer synthesis is possible in the absence of PTAS (under certain conditions), the growth area and uniformity are significantly improved when PTAS is utilized.

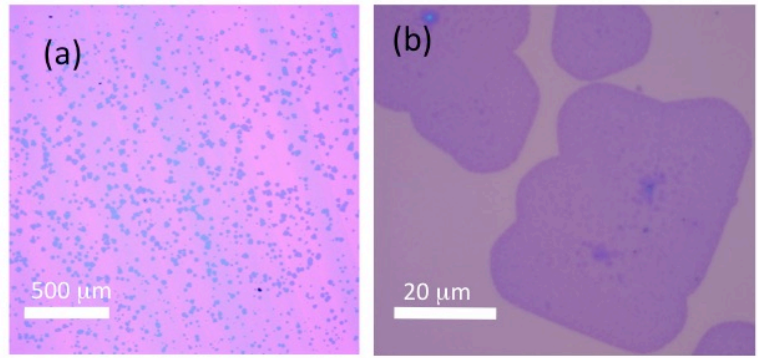

Figure S1: Optical images of WS$_2$ synthesized on the **upstream** substrate *without* the aid of PTAS seeds. (a) For synthesis conditions identical to recipe C, only isolated island growth occurs. (b) The high-magnification image shows monolayer growth in rounded-triangle shapes with a considerable amount of multilayer growth (darker purple contrast).

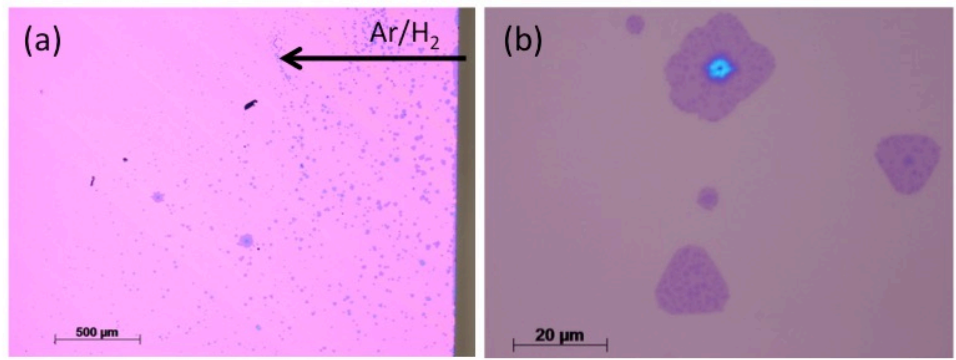

Figure S2: Optical images of WS$_2$ synthesized on the **downstream** substrate *without* the aid of PTAS seeds. (a) For synthesis conditions identical to recipe C, isolated island growth is observed within a few mm of the upstream edge. The flow direction is indicated for clarity. (b) Rounded shapes are evident in the high-magnification image, as well as multilayer WS$_2$ growth (darker purple and light blue areas).